\RequirePackage{lineno} 
\documentclass[11pt]{article}
\usepackage{graphicx}
\usepackage{bm}
\usepackage{amssymb}   
\usepackage{psfig}   

\newcommand{\bvt}{\begin{verbatim}}
\newcommand{\evt}{\end{verbatim}}

\newcommand{\as}{\alpha_s}

\newcommand{\asmz}{\alpha_s(M_Z)}

\newcommand{\pythia}{{\sc pythia}}
\newcommand{\herwig}{{\sc herwig}}
\newcommand{\djangoh}{{\sc djangoh}}
\newcommand{\rapgap}{{\sc rapgap}}
\newcommand{\ariadne}{{\sc ariadne}}
\newcommand{\lepto}{{\sc lepto}}

\newcommand{\ord}{{\cal O}}
\newcommand{\ppbar}{p{\bar{p}}}

\begin{document}

\date{September 6, 2011}

\title{\boldmath \bf 
Theory-Data Comparisons for Jet Measurements in Hadron-Induced Processes
}
\author{
The fastNLO Collaboration \\ 
\phantom{aaaaaaaaaa aaaaaaaaaaa aaaaaaaaa aaaaaaaaaaa}\\
M. Wobisch \\Louisiana Tech University, Ruston, Louisiana, USA \\
\and
D. Britzger \\  
\phantom{aaaaaaaaaa} DESY, Hamburg, Germany \phantom{aaaaaaaaaa} \\
%
\and
T. Kluge \\   University of Liverpool, U.K. \\
\and
K. Rabbertz, F. Stober \\
 \phantom{aaaaaaaaaa} KIT, Karlsruhe, Germany \phantom{aaaaaaaaaa} 
}

\maketitle

\vskip-11cm
\noindent
\phantom{.}\hfill \mbox{DESY 11-150}\\
\phantom{.}\hfill \mbox{FERMILAB-PUB-11-418-PPD}\\

\vskip4mm
\noindent

\vskip10.1cm

\begin{abstract} \noindent
We present a comprehensive overview of theory-data comparisons
for inclusive jet production.
Theory predictions are derived for recent
parton distribution functions and compared with 
jet data from different hadron-induced processes at 
various center-of-mass energies $\sqrt{s}$.
The comparisons are presented as a function of jet transverse
momentum $p_T$ or, alternatively, 
of the scaling variable $x_T = 2p_T/\sqrt{s}$.
\end{abstract}

\section{Introduction}

In this article we provide a comprehensive overview
of theory-data comparisons for 
inclusive jet cross section measurements made in hadron-induced processes
between 1993 and today.
The comparisons are made for the central region in (pseudo-) rapidity.
Data sets include inclusive jet cross sections measured in
hadron-hadron collisions at center-of-mass energies $\sqrt{s}$
between $200\,$GeV and $7\,$TeV 
and in deeply inelastic scattering (DIS)
at $\sqrt{s} = 300\,$GeV and $318\,$GeV. 
The theory calculations are computed consistently for all data sets
using recent parton distribution functions (PDFs)
and common choices for the strong coupling constant $\asmz$
and the renormalization and factorization scales.
Ratios of data and theory are presented as a function 
of jet transverse momentum\footnote{Some of the measurements 
discussed in this article have been made 
as a function of jet transverse energy $E_T$.
Throughout this article we use $p_T$ to refer 
to both, either the jet transverse momentum 
or the jet transverse energy, as appropriate.
In hadron-hadron collisions, ``transverse'' refers to the momentum
or energy component perpendicular to the hadron-hadron beam axis.
In DIS, ``transverse'' refers to the momentum component
perpendicular to the photon-proton axis 
(which is the $z$-axis in the Breit frame).}
$p_T$ or, alternatively, as a function of the
scaling variable $x_T = 2p_T/\sqrt{s}$.

This article is intended as a persistent repository for the
newest theory-data comparisons possible in jet production in
hadron-induced
processes and will be updated whenever new relevant data
and/or theory results are published.

In section 2 we discuss how the theory results are obtained
while details of the data sets are described in section 3.
Comparisons of theory and data are presented in section 4.

\section{\label{sec:theory}Theory Predictions}

Conventionally, experimental measurements are corrected 
for instrumental effects and the results are provided at the 
``particle level''~\cite{Buttar:2008jx}.
Corresponding theory predictions for jet observables
${O}_{\rm theory}$
are therefore obtained as the product of 
a perturbative QCD (pQCD) result, ${ O}_{\rm pQCD}$,
multiplied by a correction factor for non-perturbative effects,
$c_{\rm non-pert}$
\begin{equation}
 { O}_{\rm theory} =  
   { O}_{\rm pQCD}
    \cdot  c_{{\rm non-pert}}       \, .
\label{eq:allQCD}
\end{equation}
For inclusive jet cross sections in hadron-hadron collision and in DIS, 
the fixed-order perturbative expansion is known to next-to-leading order (NLO)
in the strong coupling constant $\as$, which is of order $\ord(\as^2)$ for DIS
and $\ord(\as^3)$ for hadron-hadron collisions.
For inclusive jet cross sections in hadron-hadron collisions, additional
$\ord(\as^4)$ corrections from threshold corrections have been 
computed~\cite{Kidonakis:2000gi}.
These results are used for the calculations in this article.
Renormalization and factorization scales, $\mu_r$ and $\mu_f$,
are set to $\mu_r = \mu_f = p_T$, where $p_T$ is the
transverse momentum (or, where appropriate, transverse energy)
of the individual jet.
In all cases $\mu_r$ and $\mu_f$ are larger than the b-quark mass, 
and thus all calculations are made for five active flavors ($n_f=5$). 
The proton PDFs are taken from the 
PDF parametrizations MSTW2008~\cite{Martin:2009iq},
CT10~\cite{Lai:2010vv}, 
NNPDF2.1~\cite{Ball:2011mu},
or HERAPDF1.5~\cite{herapdf15nlo, herapdf15nnlo}.
All of these parametrizations
are available in NLO and NNLO approximations 
(except for CT10 which are only available in NLO)
as well as for a series of $\asmz$ values.
In each case, we use those PDF sets which were obtained
for $\asmz = 0.118$ which is closest
to the current world average value~\cite{Bethke:2009jm}.
The same value of $\asmz$ is also employed in the matrix elements.
The NLO PDF parametrizations are used for
the comparisons with DIS data, together with the NLO matrix elements.
The NNLO versions (where available) are used for the comparisons with 
data from hadron-hadron collisions, together with NLO matrix elements
plus $\ord(\as^4)$ contributions from threshold corrections.
Correspondingly, in these comparisons $\as$ is evolved using 
the numerical solutions of
the 2-loop and 3-loop approximations of the renormalization group equation 
for DIS and hadron-hadron collisions, respectively.
All pQCD contributions are computed within the framework of
fastNLO~\cite{Kluge:2006xs}
using NLOJET++\cite{Nagy:2003tz,Nagy:2001fj} for the LO and NLO contributions
and code from the authors of Ref.~\cite{Kidonakis:2000gi}
for the threshold corrections.
The anti-$k_T$ jet algorithm is taken from the implementation in
{\sc FastJet}~\cite{Cacciari:2005hq}.

The factor $c_{\rm non-pert}$ includes non-perturbative corrections 
due to hadronization and the underlying event.
These corrections are usually obtained using Monte Carlo event generators
and the values are published together with the measurement results.
Whenever this was not the case,  we have computed the correction factors 
ourselves using \pythia\ 6.4~\cite{pythia} with tune~A~\cite{tunea}.
For DIS data at high $Q^2$, underlying event corrections are assumed
to be negligible.
In those cases, the pQCD results are only corrected for hadronization effects.

\bigskip

For some of the data sets, additional theoretical effects have already
been folded into the published results.
This is taken into account as follows:  

\begin{itemize}

\item
For some $\ppbar$ data sets (as mentioned in section 3), 
corrections for the underlying event have already been applied to the data.
In the corresponding calculations, pQCD theory is only corrected for 
hadronization effects.

\item
Observables in DIS are usually corrected for higher order QED effects
which may either include corrections for the running of the QED coupling
$\alpha$, or not.
The former is the standard choice by the ZEUS Collaboration, 
and therefore the corresponding pQCD calculations
are made for a fixed value of $\alpha = 1/137$.
Jet measurement results from the H1 Collaboration are not corrected
for the running of $\alpha$, and therefore the corresponding
pQCD calculations include a running value of $\alpha(Q)$.

\end{itemize}

\section{\label{sec:data}Data Sets}

In this section, we give a brief overview on the data sets
which are used in the theory-data comparisons.
The phase space requirements are given as stated in the cited publications
which include further details regarding the exact definition 
of the variables.
The recombination schemes used in the jet algorithms are usually the
 $E$-scheme~\cite{run2cone}, 
or the $E_T$-scheme (frequently referred to 
as ``Snowmass convention'')~\cite{Huth:1990mi}.
In some cases other recombination schemes have been used.
The following measurements of the inclusive jet cross section
as a function of $p_T$ are used in the comparisons. 
All data are corrected to the particle level as defined 
in Ref.~\cite{Buttar:2008jx}, except where mentioned otherwise.

\begin{enumerate}

\item Proton-proton scattering at $\sqrt{s} = 7\,$TeV at the LHC
\begin{enumerate}

\item CMS Collaboration (2011) 
      ${\cal L_{\rm int}}=34\,$pb$^{-1}$~\cite{:2011me}  \\
Measurement of
$d^2\sigma/dp_T dy$ for $18<p_T<1100\,$GeV
      in six rapidity regions $|y|<3.0$ 
using the
anti-$k_T$ jet algorithm with $R=0.5$ in the $E$-scheme. 
In our comparisons we display the results for $|y|<0.5$.  
Non-perturbative corrections are determined 
as the average of \pythia\ 6.422 with tune D6T~\cite{Field:2008zz}
and \herwig++ 2.4.2~\cite{Bahr:2008pv} with the default tune
of \herwig++ 2.3
(as detailed in Ref.~\cite{Rabbertz:2011}).

\item ATLAS Collaboration (2010) 
      ${\cal L_{\rm int}}=17\,$nb$^{-1}$~\cite{:2010wv} \\
Measurement of
$d^2\sigma/dp_T dy$  for $60<p_T<500\,$GeV
     in five rapidity regions $|y|<2.8$ 
using the 
anti-$k_T$ jet algorithm with $R=0.4$ in the $E$-scheme. 
In our comparisons we display the results for $0<|y|<0.3$.   
Non-perturbative corrections are determined 
using \pythia\ 6 with tune MC09.

\item ATLAS Collaboration (2010) 
      ${\cal L_{\rm int}}=17\,$nb$^{-1}$~\cite{:2010wv}  \\
The same as in 1.~(b), but for $R=0.6$.

\end{enumerate}

\item Proton-anti-proton scattering at $\sqrt{s} = 1.96\,$TeV at the Fermilab Tevatron collider
\begin{enumerate}

\item CDF Collaboration (2008) 
      ${\cal L_{\rm int}}=1.13\,$fb$^{-1}$~\cite{Aaltonen:2008eq} \\
Measurement of
$d^2\sigma/dp_T dy$  for $62<p_T<700\,$GeV
     in five rapidity regions $|y|<2.1$ 
using the
Run II iterative midpoint cone jet algorithm (CDF version) with $R=0.7$ 
in the $E$-scheme. 
In our comparisons we display the results for $0.1<|y|<0.7$.  
Non-perturbative corrections are determined 
using \pythia\ with tune A.

\item CDF Collaboration (2007) 
      ${\cal L_{\rm int}}=1.0\,$fb$^{-1}$~\cite{Abulencia:2007ez} \\
Measurement of
$d^2\sigma/dp_T dy$  for $52<p_T<700\,$GeV
     in five rapidity regions $|y|<2.1$ 
using the 
inclusive $k_T$ jet algorithm  with $R=0.7$ in the $E$-scheme. 
In our comparisons we display the results for $0.1<|y|<0.7$.
Non-perturbative corrections are determined 
using \pythia\ with tune A.

\item D\O\ Collaboration (2008) 
      ${\cal L_{\rm int}}=0.7\,$fb$^{-1}$~\cite{:2008hua}   \\
Measurement of
$d^2\sigma/dp_T dy$  for $50<p_T<600\,$GeV
     in six rapidity regions $|y|<2.4$
using the 
Run II iterative midpoint cone jet algorithm (D\O\ version) with $R=0.7$ 
in the $E$-scheme. 
In our comparisons we display the results for $|y|<0.4$.  
Non-perturbative corrections are determined 
using \pythia\ with tune QW~\cite{Group:2006rt}.

\end{enumerate}

\item Proton-anti-proton scattering at $\sqrt{s} = 1.8\,$TeV at the Fermilab Tevatron collider
\begin{enumerate}

\item CDF Collaboration (2001) 
      ${\cal L_{\rm int}}=10\,$pb$^{-1}$~\cite{Affolder:2001fa} \\
Measurement of
$d^2\sigma/dE_T d\eta$  for $40<E_T<465\,$GeV
     in the pseudorapidity range $0.1<|\eta|<0.7$ 
using the
JETCLU iterative cone jet algorithm with $R=0.7$  
    in a special recombination scheme
    described in equations (3-9) in~\cite{Affolder:2001fa}. 
Corrections for the effects due to the underlying event 
have been estimated using ambient energy measured in 
minimum bias events and applied to the data. 
Hadronization corrections have not been published.

\item D\O\ Collaboration (2001) 
      ${\cal L_{\rm int}}=95\,$pb$^{-1}$~\cite{Abbott:2000ew}  \\
Measurement of
$d^2\sigma/dE_T d\eta$  for $60<E_T<500\,$GeV
     in five pseudorapidity regions $|\eta|<3.0$ 
using the 
D\O\ Run I iterative cone jet algorithm with $R=0.7$ 
  in a special recombination scheme, but with additional corrections 
  of jet $\eta$ to account for the differences 
  to the $E_T$-scheme~\cite{Babukhadia}.
In our comparisons we display the results for $|\eta|<0.5$.
Corrections for the effects due to the underlying event 
have been estimated using low luminosity minimum bias
data~\cite{Abbott:1998xw} and applied to the data.
Hadronization corrections have not been published.

\end{enumerate}

\item Proton-anti-proton scattering at $\sqrt{s} = 546$, $630\,$GeV at the Fermilab Tevatron collider
\begin{enumerate}

\item CDF Collaboration (1993) 
     at $\sqrt{s} = 546\,$GeV,
      ${\cal L_{\rm int}}=8.58\,$nb$^{-1}$~\cite{Abe:1992bk} \\
Measurement of
$d^2\sigma/dE_T d\eta$  for $27.5<E_T<73\,$GeV
     in the pseudorapidity range $0.1<|\eta|<0.7$ 
using the
JETCLU iterative cone jet algorithm with $R=0.7$  in a special recombination scheme,
    described in equations (3-9) in~\cite{Affolder:2001fa}.
Corrections for effects due to the underlying event 
have been estimated using the observed calorimeter transverse energy 
at $90^\circ$ to the jet axis in CDF dijet events
and applied to the data. 
Hadronization corrections have not been published.

\item D\O\ Collaboration (2001) 
  $\sqrt{s} = 630\,$GeV,
     at ${\cal L_{\rm int}}=538\,$nb$^{-1}$~\cite{Abbott:2000kp}   \\
Measurement of
$d^2\sigma/dE_T d\eta$  for $21<E_T<196\,$GeV
     in the pseudorapidity region $|\eta|<0.5$ 
using the
D\O\ Run I iterative cone jet algorithm with $R=0.7$ 
  in a special recombination scheme (see~\cite{Abbott:2000kp}).
Corrections for the effects due to the underlying event 
have been estimated using Monte Carlo simulations 
of $\ppbar$ interactions and applied to the data. 
Hadronization corrections have not been published.

\end{enumerate}

\item Proton-proton scattering at $\sqrt{s} = 200\,$GeV at RHIC
\begin{enumerate}

\item STAR Collaboration (2006) 
      ${\cal L_{\rm int}}=0.30\,$pb$^{-1}$~\cite{Abelev:2006uq} \\
Measurement of
$d^2\sigma/dp_T d\eta$  for $5<p_T<49\,$GeV
     in the pseudorapidity region $0.2<|\eta|<0.8$ 
using the 
Run II iterative midpoint cone jet algorithm 
(implemented following~\cite{run2cone})
           with $R=0.7$ in the $E$-scheme. 
Non-perturbative corrections have not been published.
Those were provided to us by one of the authors~\cite{miller2006}. \\

\end{enumerate}

\item Deeply inelastic scattering at $\sqrt{s} = 318\,$GeV at HERA
\begin{enumerate}

\item H1 Collaboration (2007) 
      ${\cal L_{\rm int}}=65.4\,$pb$^{-1}$~\cite{:2007pb} \\
Measurement of
$d^2\sigma/dE_T dQ^2$  for $7<E_T<50\,$GeV
    and $150<Q^2<15000\,$GeV$^2$
     at an inelasticity $0.2<y< 0.7$ 
     in the pseudorapidity region $-1<\eta_{\rm lab}<2.5$
using the
inclusive $k_T$ jet algorithm in the Breit frame with $R=1$ 
       in the $E_T$-scheme.
In these comparisons we are using the data 
$150<Q^2<5000\,$GeV$^2$ for which contributions from $Z$ exchange
are negligible. 
The data have been corrected to the particle level 
and for QED radiative corrections, 
but not for the running of $\alpha$. 
Hadronization corrections have been determined 
as the average value using 
\djangoh~\cite{Charchula:1994kf} and \rapgap~\cite{Jung:1993gf}.

\item ZEUS Collaboration (2007) 
      ${\cal L_{\rm int}}=81.7\,$pb$^{-1}$~\cite{Chekanov:2006xr} \\
Measurement of
$d^2\sigma/dE_T dQ^2$  for $8<E_T<100\,$GeV
    and $Q^2>125\,$GeV$^2$
     with $|\cos \gamma_h| < 0.65$
     in the pseudorapidity region $-2<\eta_{\rm Breit}<1.5$ 
using the
inclusive $k_T$ jet algorithm in the Breit frame with $R=1$ 
       in the $E_T$-scheme.
In these comparisons we are using the data 
$125<Q^2<5000\,$GeV$^2$ for which contributions from $Z$ exchange
are negligible. 
The data have been corrected to particle level 
and for QED radiative corrections, 
including the running of $\alpha$. 
Hadronization corrections have been determined 
as the average value using 
\ariadne~\cite{Lonnblad:1992tz,Lonnblad:1994wk}
and \lepto~\cite{Ingelman:1996mq}.

\end{enumerate}

\item Deeply inelastic scattering at $\sqrt{s} = 300\,$GeV at HERA
\begin{enumerate}

\item H1 Collaboration (2001) 
      ${\cal L_{\rm int}}=33\,$pb$^{-1}$~\cite{Adloff:2000tq} \\
Measurement of
$d^2\sigma/dE_T dQ^2$  for $7<E_T<50\,$GeV
    and $150<Q^2<5000\,$GeV$^2$
     at $0.2<y< 0.6$ 
     in the pseudorapidity region $-1<\eta_{\rm lab}<2.5$ 
using the
inclusive $k_T$ jet algorithm in the Breit frame with $R=1$ 
       in the $E_T$-scheme.
The data have been corrected to particle level 
and for QED radiative corrections, 
but not for the running of $\alpha$. 
Hadronization corrections have been determined 
as the average value from \herwig~\cite{Marchesini:1991ch}, 
\lepto, and \ariadne.

\item ZEUS Collaboration (2002) 
      ${\cal L_{\rm int}}=10\,$pb$^{-1}$~\cite{Chekanov:2002be} \\
Measurement of
$d^2\sigma/dE_T dQ^2$  for $8<E_T<100\,$GeV
    and $Q^2>125\,$GeV$^2$
     with $-0.7 <\cos \gamma_h < 0.5$
     in the pseudorapidity region $-2<\eta_{\rm Breit}<1.8$ 
using the
inclusive $k_T$ jet algorithm in the Breit frame with $R=1$ 
       using the $E_T$-scheme.
In these comparisons we are using the data 
$125<Q^2<5000\,$GeV$^2$ for which contributions from $Z$ exchange
are negligible. 
The data have been corrected to particle level 
and for QED radiative corrections, 
including the running of $\alpha$. 
Hadronization corrections have been determined 
as the average value using \ariadne, \lepto, and \herwig.

\end{enumerate}

\end{enumerate}

\newpage
\section{Results}

\begin{figure}
  \centerline{
\includegraphics[scale=0.7]{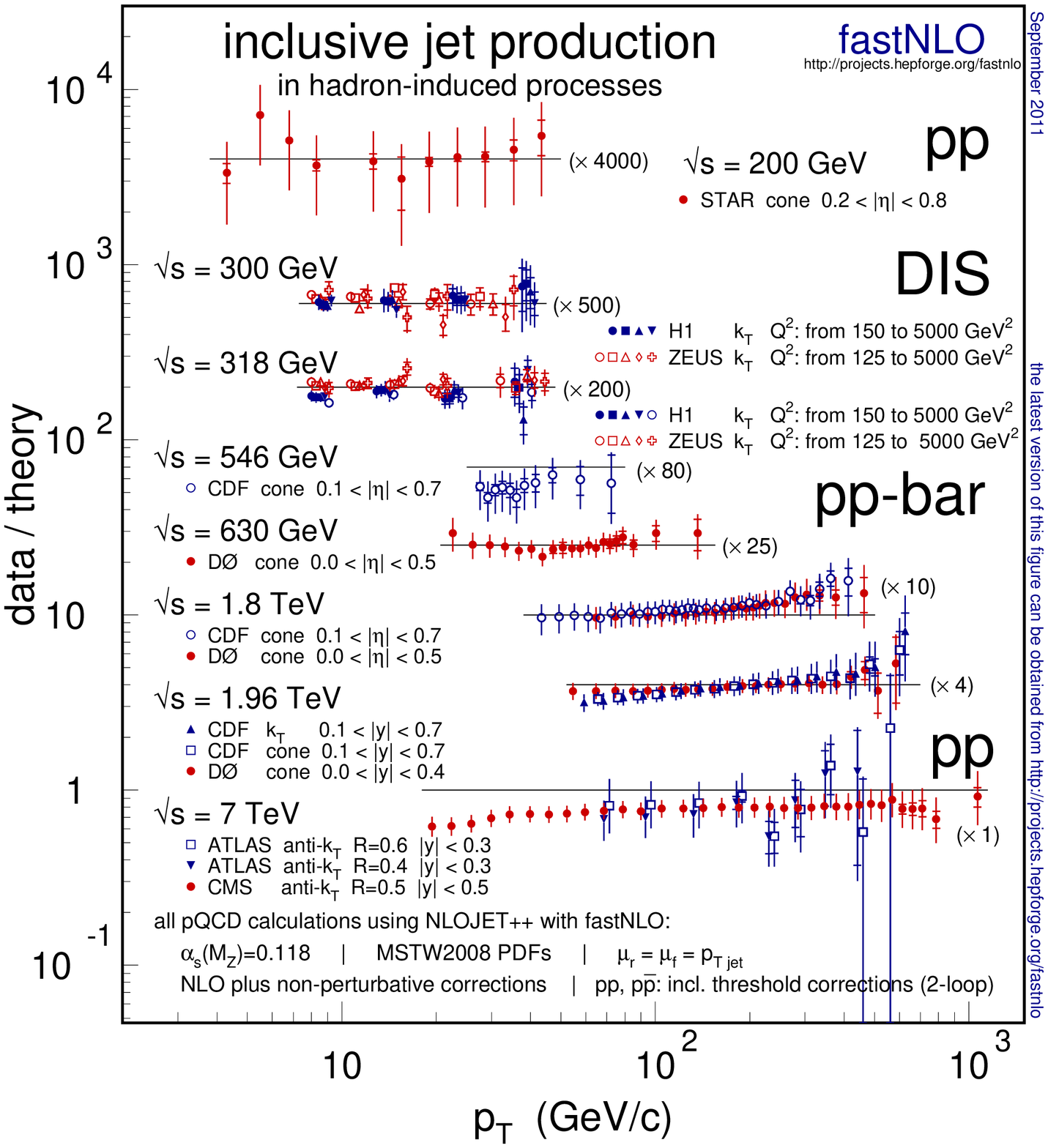}%
}
\caption{\label{fig:incljets}
Ratios of data and theory for inclusive jet cross sections
measured in hadron-hadron collisions and in deeply inelastic scattering
at different center-of-mass energies.
The ratios are shown as a function of jet transverse momentum $p_T$.
The theory results are computed for MSTW2008 PDFs.
}
\end{figure}

The ratios of data and theory for all results
obtained in hadron-hadron collisions 
and in DIS at large four-momentum transfer squared $Q^2$,
are shown in Fig.~\ref{fig:incljets} as a function
of jet $p_T$.
For DIS, the variable $p_T$ corresponds to the jet transverse 
energy defined in the Breit frame of reference.
The uncertainty bars indicate the total experimental uncertainties,
defined as the quadratic sum of statistical and systematic uncertainties.
For purposes of visibility, the ratios for different 
center-of-mass energies are multiplied by factors as given in the figures,
and the results are sorted from top to bottom
in the order of increasing center-of-mass energy.

\begin{figure}
  \centerline{
\includegraphics[scale=0.7]{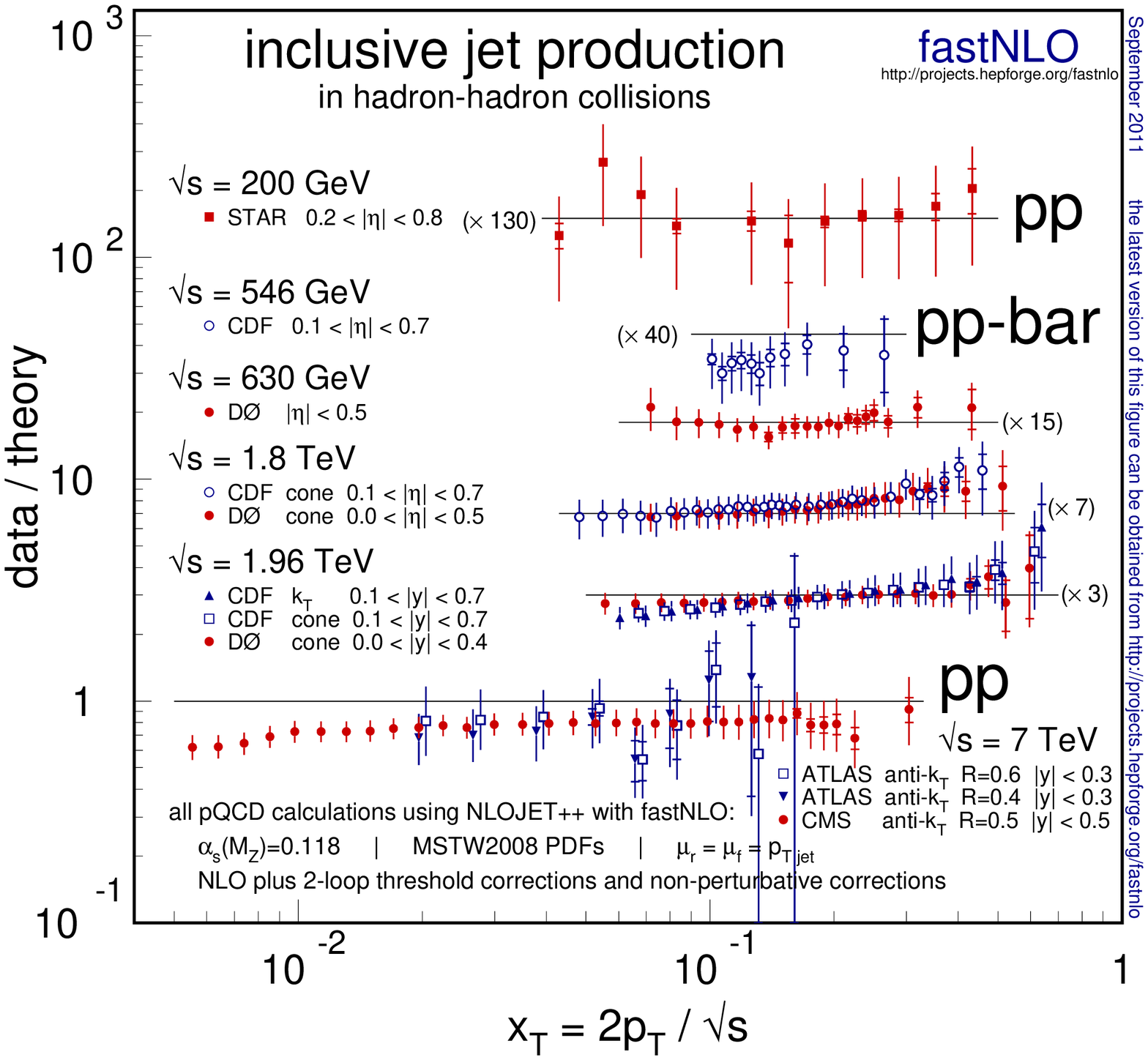}%
}
\caption{\label{fig:ppjetsxt}
Ratios of data and theory for inclusive jet cross sections
measured in hadron-hadron collisions
at different center-of-mass energies.
The ratios are shown as a function of the scaling variable
$x_T = 2p_T/\sqrt{s}$.
The theory results are computed for MSTW2008 PDFs.
}
\end{figure}

Figure~\ref{fig:ppjetsxt} shows the data points
from hadron-hadron collisions, this time as a function 
of $x_T = 2p_T/\sqrt{s}$.
For central jet production in hadron-hadron collisions, 
the variable $x_T$ is closely related to the hadron 
momentum fractions $x_1$ and $x_2$, carried by the partons.
In exclusive dijet production with
rapidities $y_1=y_2=0$, the relation $x_T= x_1 = x_2$ holds.

The comparison of inclusive jet cross section data at central 
(pseudo-) rapidities as a function of $x_T$
allows to infer the sensitivity of PDF determinations 
to the measured data.

In general, we observe good agreement between theory and data.
The data from some measurements 
(CDF at $\sqrt{s}=546\,$GeV~\cite{Abe:1992bk} and 
ATLAS and CMS at $\sqrt{s}=7\,$TeV~\cite{:2010wv,:2011me})
have a tendency of being systematically below 
the theory predictions.
However, the CDF and ATLAS data have relatively large uncertainties,
and it was discussed in Refs.~\cite{:2011me,Rabbertz:2011} that
within experimental and theoretical uncertainties, 
the CMS data are in agreement with theory.
Differences in the shapes between theory and data as a function
of $p_T$ for the jet data at $\sqrt{s}=1.8$ and $1.96\,$TeV
can be explained by correlated experimental uncertainties,
as studied in recent global 
PDF analyses~\cite{Martin:2009iq,Lai:2010vv,Ball:2011mu}.

\begin{figure}
  \centerline{
\includegraphics[scale=0.7]{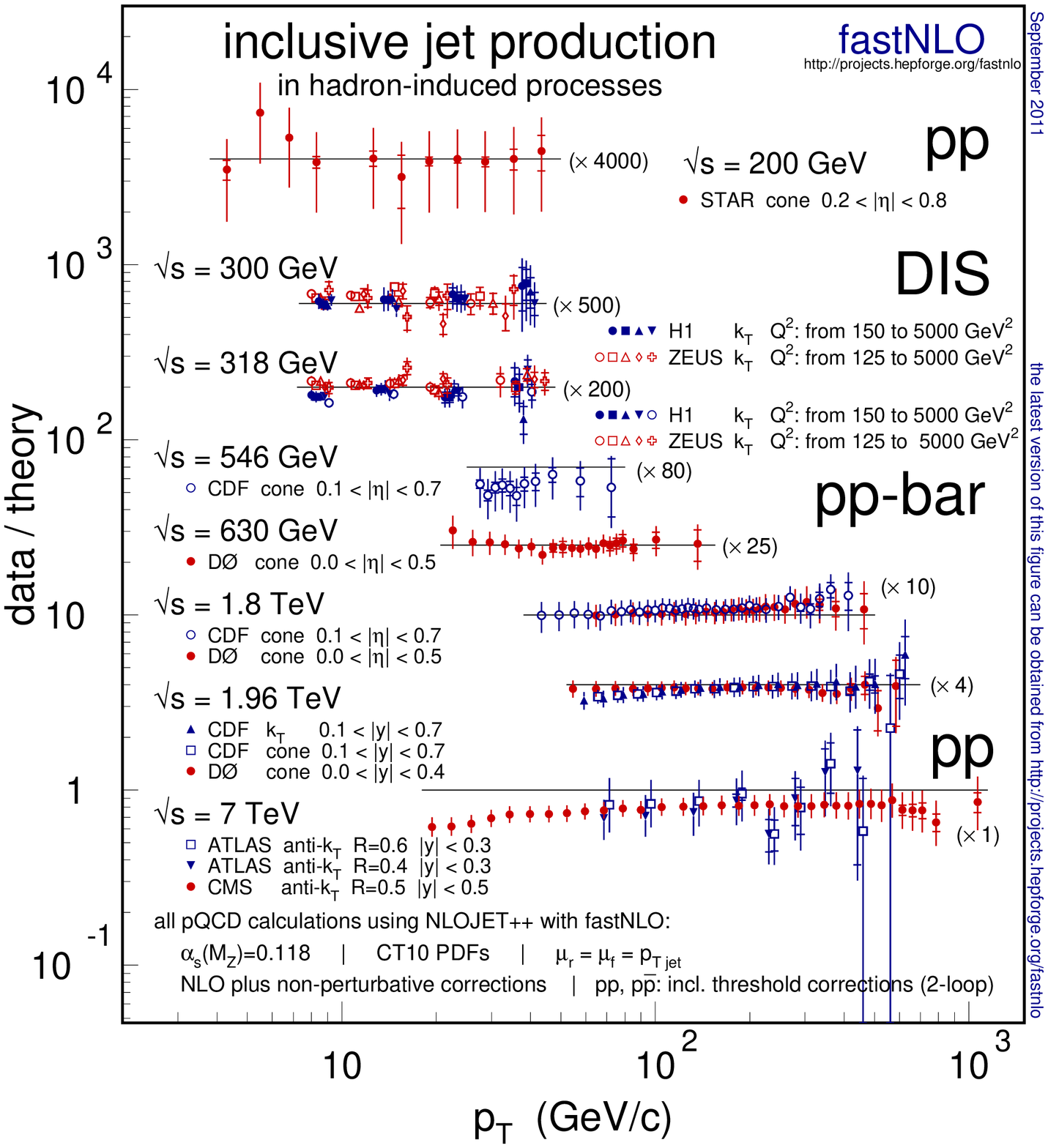}%
}
\caption{\label{fig:incljets-ct10}
Ratios of data and theory for inclusive jet cross sections
measured in hadron-hadron collisions and in deeply inelastic scattering
at different center-of-mass energies.
The ratios are shown as a function of jet transverse momentum $p_T$.
The theory results are computed for CT10 PDFs.}
\end{figure}

\begin{figure}
  \centerline{
\includegraphics[scale=0.7]{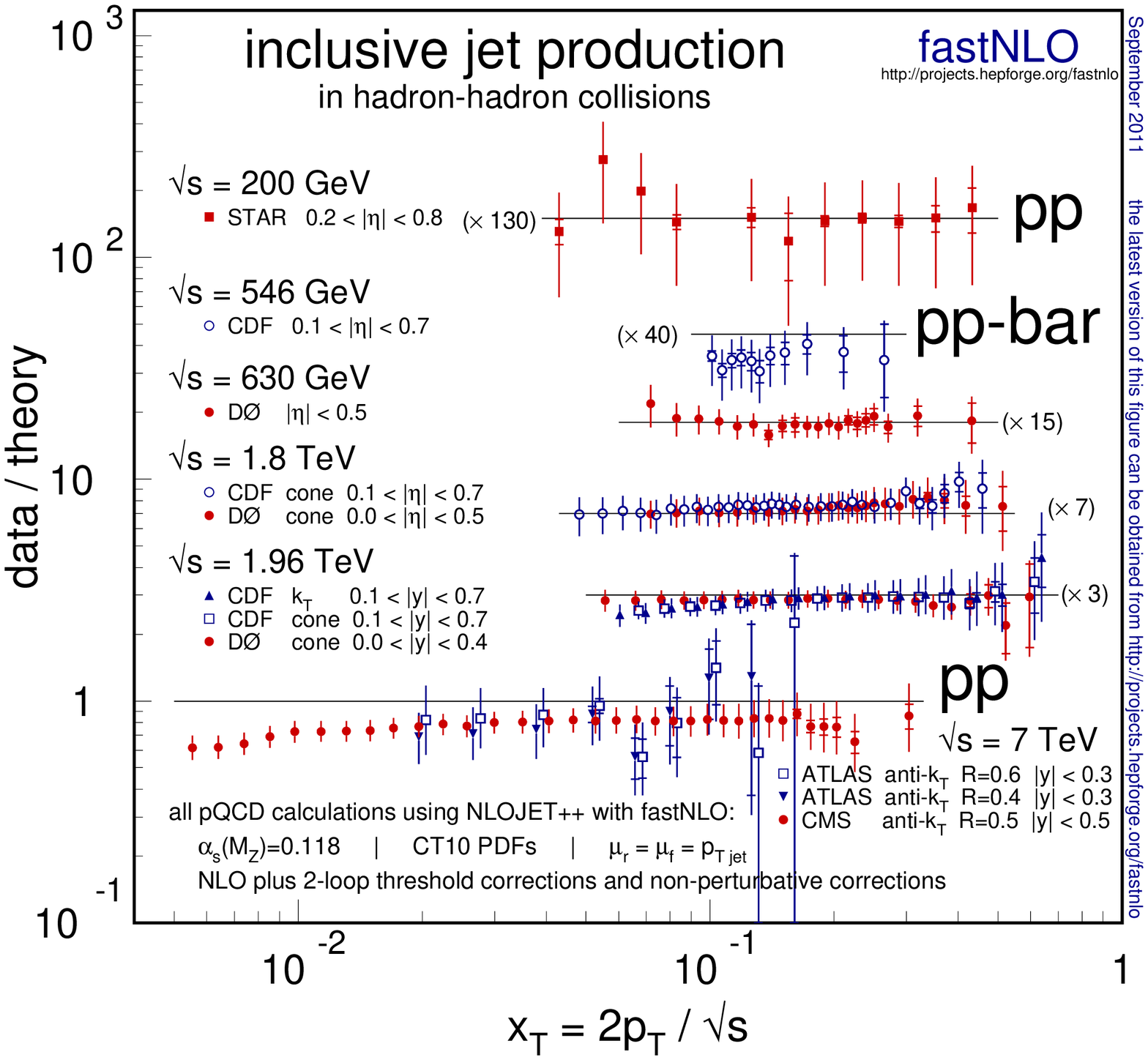}%
}
\caption{\label{fig:ppjetsxt-ct10}
Ratios of data and theory for inclusive jet cross sections
measured in hadron-hadron collisions
at different center-of-mass energies.
The ratios are shown as a function of the scaling variable
$x_T = 2p_T/\sqrt{s}$.
The theory results are computed for CT10 PDFs.}
\end{figure}

\begin{figure}
  \centerline{
\includegraphics[scale=0.7]{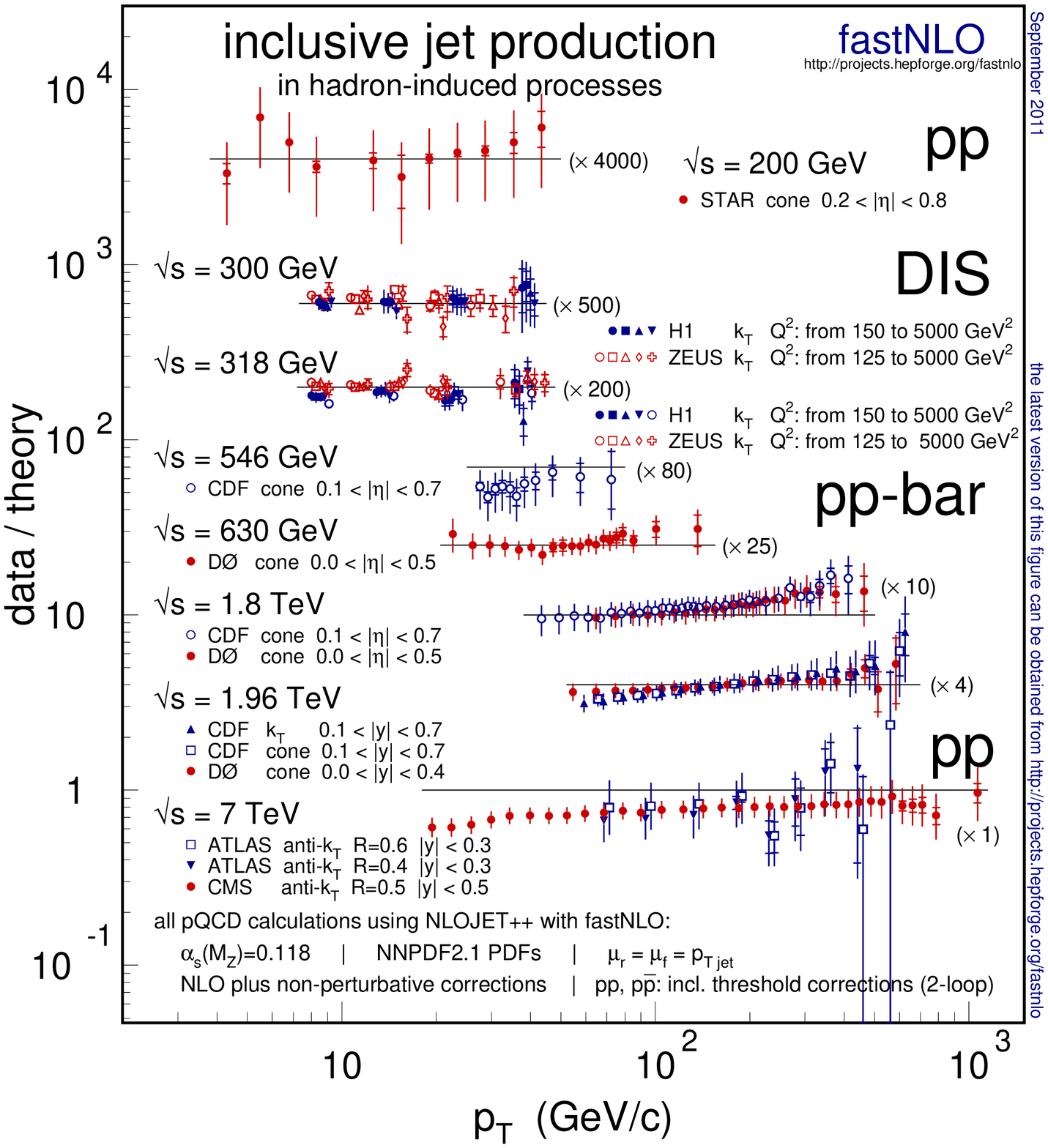}%
}
\caption{\label{fig:incljets-nnpdf}
Ratios of data and theory for inclusive jet cross sections
measured in hadron-hadron collisions and in deeply inelastic scattering
at different center-of-mass energies.
The ratios are shown as a function of jet transverse momentum $p_T$.
The theory results are computed for NNPDF2.1 PDFs.}
\end{figure}

\begin{figure}
  \centerline{
\includegraphics[scale=0.7]{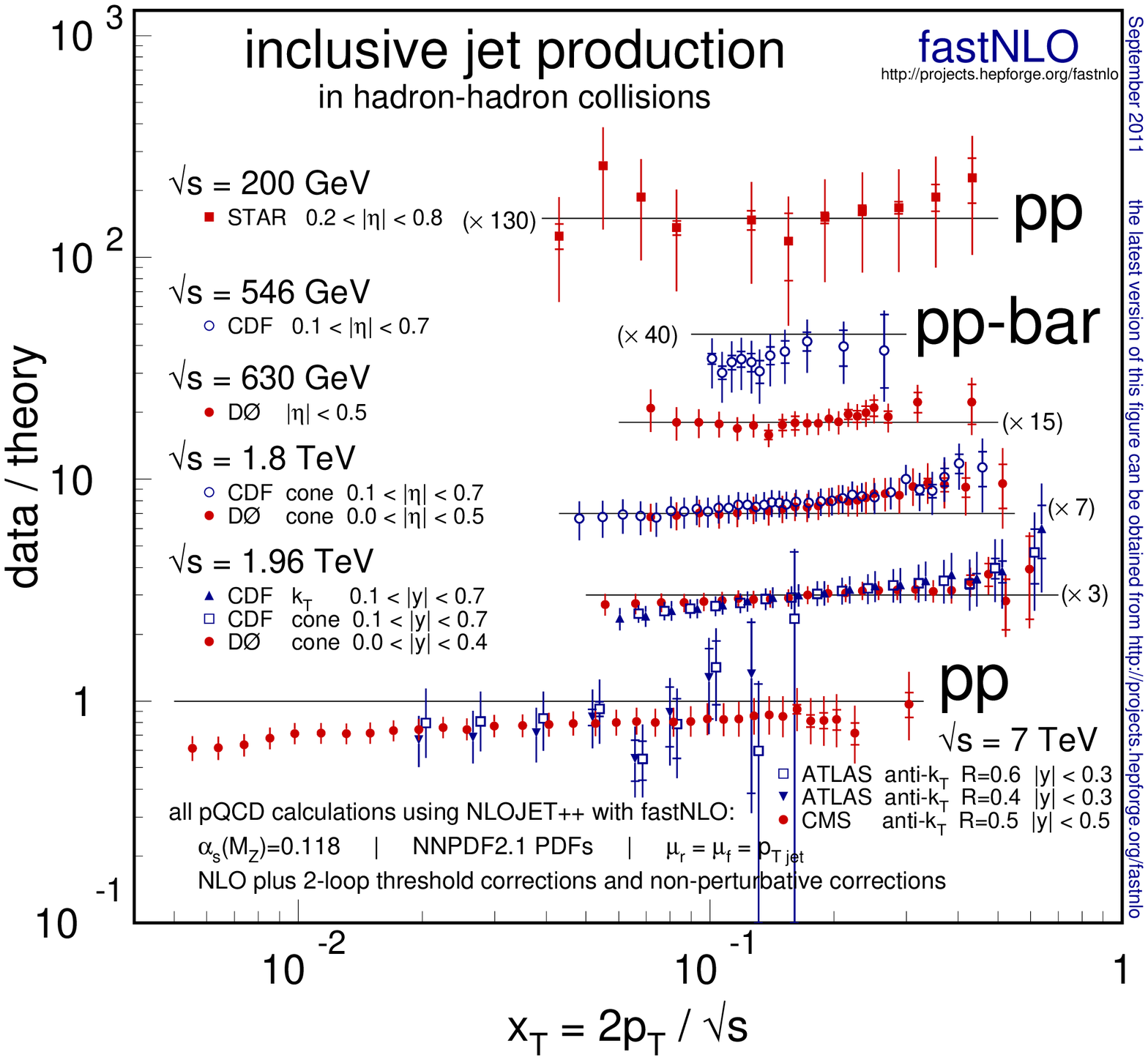}%
}
\caption{\label{fig:ppjetsxt-nnpdf}
Ratios of data and theory for inclusive jet cross sections
measured in hadron-hadron collisions
at different center-of-mass energies.
The ratios are shown as a function of the scaling variable
$x_T = 2p_T/\sqrt{s}$.
The theory results are computed for NNPDF2.1 PDFs.}
\end{figure}

\begin{figure}
  \centerline{
\includegraphics[scale=0.7]{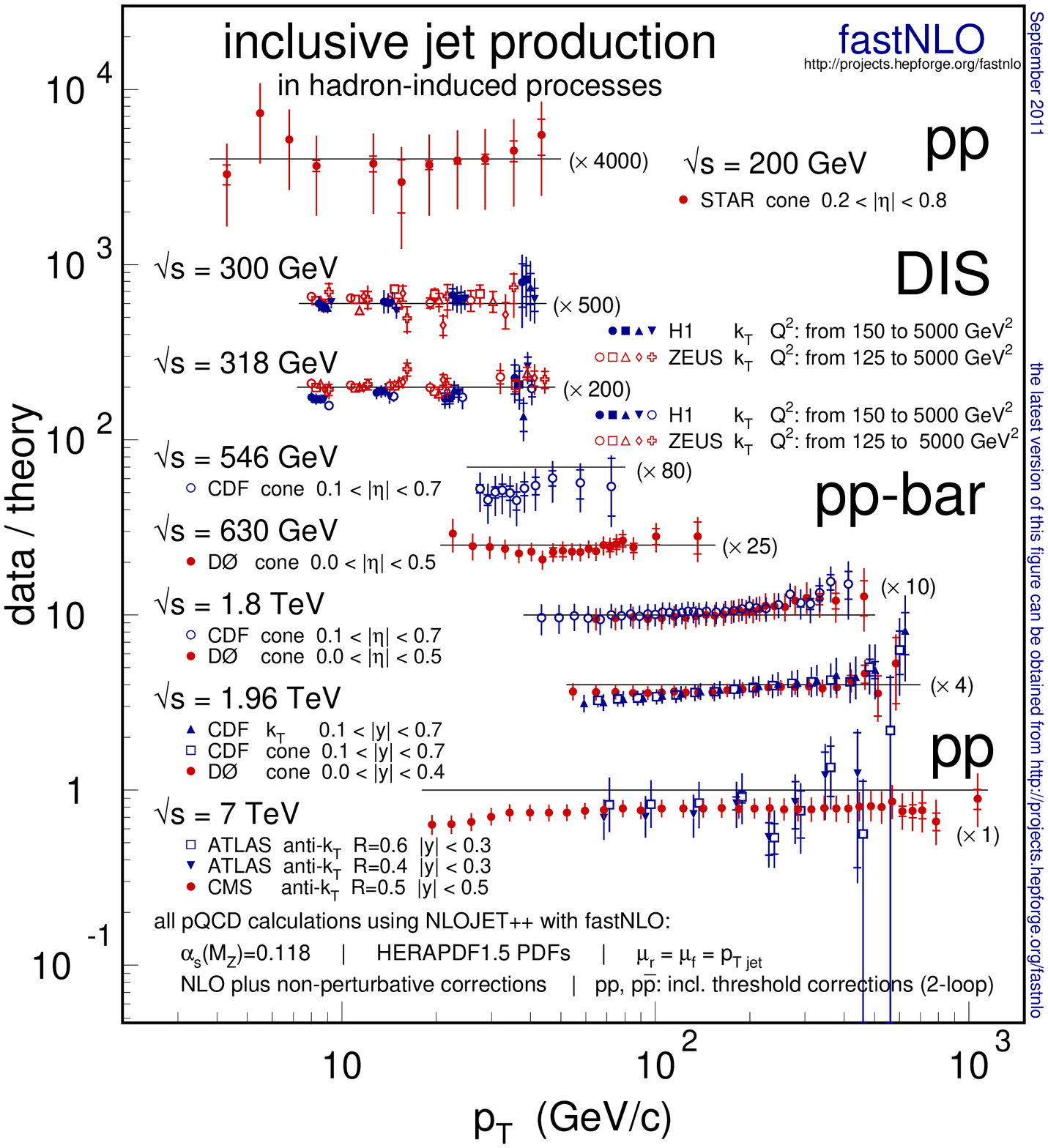}%
}
\caption{\label{fig:incljets-hera}
Ratios of data and theory for inclusive jet cross sections
measured in hadron-hadron collisions and in deeply inelastic scattering
at different center-of-mass energies.
The ratios are shown as a function of jet transverse momentum $p_T$.
The theory results are computed for HERAPDF1.5 PDFs.
}
\end{figure}

\begin{figure}
  \centerline{
\includegraphics[scale=0.7]{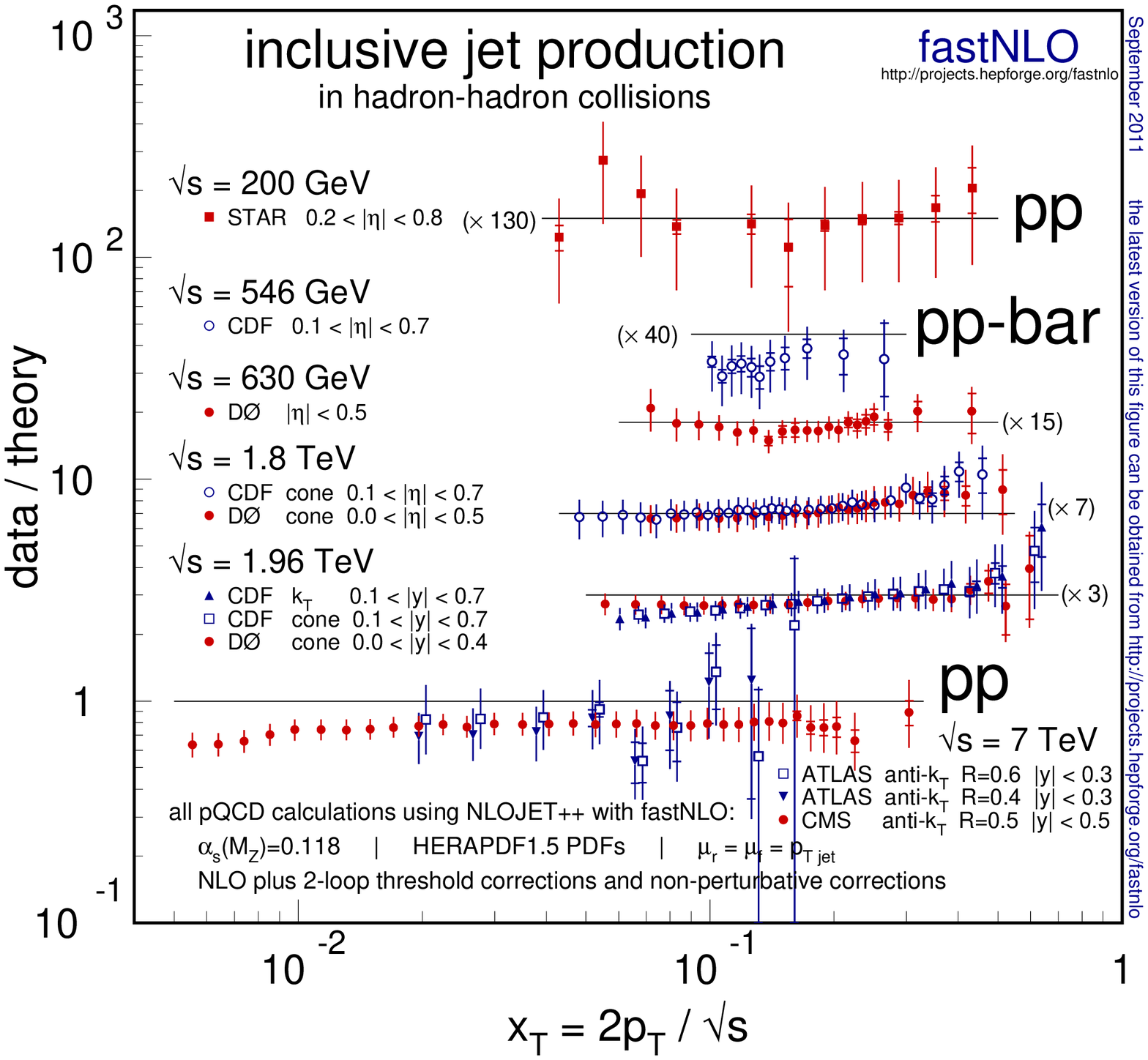}%
}
\caption{\label{fig:ppjetsxt-hera}
Ratios of data and theory for inclusive jet cross sections
measured in hadron-hadron collisions
at different center-of-mass energies.
The ratios are shown as a function of the scaling variable
$x_T = 2p_T/\sqrt{s}$.
The theory results are computed for HERAPDF1.5 PDFs.
}
\end{figure}

The theory-data comparisons which are presented 
in Figs.~\ref{fig:incljets} and~\ref{fig:ppjetsxt}
for MSTW2008 PDFs are also repeated 
for the PDF parametrizations CT10, NNPDF2.1, and HERAPDF1.5.
These results are displayed in
Figs.~\ref{fig:incljets-ct10} -- \ref{fig:ppjetsxt-hera}.
In most cases we observe a very similar agreement as for MSTW2008 PDFs.

From Fig.~\ref{fig:incljets}
it is visible that the new LHC jet data have started to go beyond 
the $p_T$ reach of the Fermilab Tevatron experiments CDF and D\O.
Fig.~\ref{fig:ppjetsxt} shows that the new CMS measurement
provides the first jet results which probe the PDFs down 
to $x = 5\cdot10^{-3}$.
In this $x$-region, the PDFs are already constrained by
data from inclusive DIS experiments.
Therefore the comparison of LHC jet data with theory predictions in this
overlap region will help to understand
uncertainties not related to PDFs.
It is also visible from Fig.~\ref{fig:ppjetsxt} that 
currently the Tevatron jet data still have the highest reach in $x$.



\end{document}